# Quantitative strain-field measurement of 1:1 B-site cation ordered domains and antiphase boundaries in Pb(Sc$_{1/2}$Ta$_{1/2}$)O$_3$ ceramics by high-resolution transmission electron microscopy

Cheuk W. Tai[a)] and Y. Lereah

Department of Physical Electronics, School of Electrical Engineering,

The Iby and Aladar Fleischman Faculty of Engineering,

Tel Aviv University, Ramat Aviv 69978, Israel

**Abstract**

Quantitative strain measurements of the 1:1 B-site cation ordered domains, antiphase boundaries and dislocations in a highly ordered Pb(Sc$_{1/2}$Ta$_{1/2}$)O$_3$ ceramic have been carried out by high-resolution transmission electron microscopy and geometric phase analysis. A phase shift of π between two adjacent ordered domains across an antiphase boundary are determined unambiguously. The maximum in-plane strain and lattice rotation induced by a dislocation are 9.5% and 5.4°, respectively. In a defect-free antiphase boundary, the maximum in-plane strain and lattice rotation are 1.8% and 0.9°, respectively. The strain mainly concentrates inside the antiphase boundary.

---

[a)] Electronic mail: cheukw.tai@gmail.com

**Introduction**

Lead scandium tantalate, $Pb(Sc_{1/2}Ta_{1/2})O_3$ (PST), is a well-known order-disorder perovskite ferroelectric material. It has been studied extensively because of its pyroelectric property and order-disorder behavior.[1,2,3,4,5,6,7,8] The $Pb^{2+}$ cations occupy the A-sites of the perovskite structure, while $Sc^{3+}$ and $Ta^{5+}$ cations occupy the B-sites at the center of the oxygen octahedra. The $Sc^{3+}$ and $Ta^{5+}$ cations may occupy the B-site sites randomly or can develop 1:1 NaCl-type order on the {111} planes. The disordered or intermediate ordered PST shows the ferroelectric relaxor behavior, but becomes a normal ferroelectric below the Curie temperature in highly ordered state. The configuration of the two B-site cations strongly influences the physical properties of PST, such as electrical properties and phase transition behavior.[3,4,5,6,7,8] In contrast, the elastic properties of PST are rarely reported.[9,10] Besides polarization, the microscopic strain states and elastic energy are the important parameters in the theories of ferroelectric/relaxor.[11,12,13] Besides chemical heterogeneity, strain is considered as a source of the random fields in relaxors.[14]

The degree of 1:1 cation order (S) can be characterized by Raman spectroscopy, x-ray, neutron and electron diffraction.[15,16,17,18,19] However, the spatial and size distribution of the ordered domains in a PST grain can be obtained by dark-field imaging and high-resolution transmission electron microscopy (HRTEM).[19,20] In a Bragg-filtered HRTEM image,[21] an antiphase boundary (APB) can be distinguished from the disordered domains. A phase shift of π between the superstructure lattice fringes occurs across an APB. However, the above techniques are unable to give the quantitative structural information such as strain.

The geometric phase analysis (GPA)[22] performs the measurement of local displacement with respect to a reference region in a HRTEM image and then calculates the corresponding strain tensor $\varepsilon_{ij}$ and rotation $\omega_{xy}$. This method has already been applied to a number of quantitative HRTEM studies.[22,23,24,25] A particular set of fringes in a HRTEM image can be selected by applying a filter on its Fourier component **g** and the intensity is described as:

$$\mathbf{I_g(r)} = A_g(\mathbf{r})\exp[i\mathbf{P_g(r)}] \qquad (1)$$

where the amplitude $A_g(\mathbf{r})$ and phase $\mathbf{P_g(r)}$ are represent the contrast level and location of the fringes. The relative phase shift induced by a small displacement of the fringes $\mathbf{u(r)}$ with respect to a reference lattice can be measured:

$$\mathbf{P_g(r)} = -2\pi \mathbf{g} \cdot \mathbf{u(r)} \qquad (2)$$

The two-dimensional displacement field can be obtained by two non-collinear **g**:

$$\mathbf{u(r)} = \frac{-1}{2\pi}[\mathbf{P_{g1}(r)a_1} + \mathbf{P_{g2}(r)a_2}] \qquad (3)$$

where $\mathbf{a_1}$ and $\mathbf{a_2}$ are the lattice vector in real space corresponding to $\mathbf{g_1}$ and $\mathbf{g_2}$, respectively. The strain tensor and rotation for a small deformation is given by the derivatives of $\mathbf{u(r)}$:

$$\varepsilon_{ij} = \frac{1}{2}(\frac{\partial u_i}{\partial x_j} + \frac{\partial u_j}{\partial x_i}) \text{ and } \omega_{xy} = \frac{1}{2}(\frac{\partial u_y}{\partial x} - \frac{\partial u_x}{\partial y}) \qquad (4)$$

**Experimental**

Highly ordered PST samples (S=0.86) were fabricated using mixed-oxide route.[26,27,28] TEM samples were prepared by polishing and then Ar$^+$ ion-milling. The HRTEM study was carried out at room temperature in a 200 kV FEI Tecnai F20ST microscope ($C_s$ = 1.2 mm). The quantitative analysis of the images was done by the package, GPA Phase (HREM Research Inc.). A Lorentzian filter in radius of 0.75nm$^{-1}$ ($|\mathbf{g}_{100}|/3$) was applied. HRTEM images of a Si crystal were used to evaluate the detectivity of our microscope. In a defect-free region in Si, the fluctuation of the measured in-plane strain and rotation are ±0.5 % and ±0.2°, respectively. The images were carefully selected to avoid the influence of imaging artifacts on the GPA results, such as contrast reversals and thickness changes.[29] Image delocalization does not contribute significantly to this study. It can be estimated by:[30]

$$\Delta R = |\nabla \chi(\mathbf{g})| = \lambda |\mathbf{g}| \Delta f + C_s \lambda^3 |\mathbf{g}|^3 \quad (5)$$

where $\chi$ is the aberration function, $\lambda$ is the electron wavelength (0.0025 nm) and $\Delta f$ is the defocus (-100 nm). If we consider $\mathbf{g}_{110}$, the delocalization is ~8 Å that is less than two unit cells.

**Observations and Discussion**

Figure 1(a) is a [110] HRTEM image of two ordered domains. The domains are separated by a 3 nm wide APB. The power spectrum of the HRTEM image is shown in Fig. 1(b). It was indexed according to space group *Pm3m*. The high intensity of $\mathbf{g}_{1/2 1/2 1/2}$ indicates the presence of the highly ordered domains. In the phase images shown in Fig. 1(c) and (d), a region in the OD 1 far from the APB is selected as the reference in GPA. The value of the phase in OD 1 is zero on average and the fluctuation is less than $\pm 0.08\pi$, indicating no change of fringe spacing. The position of the APB is revealed by the discontinuity of phase. A phase jump from 0 to $\pi$ (or $-\pi$) is caused by the superstructure fringes in two adjacent ordered domains shifted relatively by half a period.

In order to characterize the local lattice deformation, the (-110) and (001) lattice fringes ($\mathbf{g}_{-110}$ and $\mathbf{g}_{001}$) are selected in the GPA. The maps of the in-plane strain fields ($\varepsilon_{xx}$, $\varepsilon_{yy}$ and $\varepsilon_{xy}$) and lattice rotation ($\omega_{xy}$) are shown in Figs. 2(a)-(d), respectively. The values of $\varepsilon_{xx}$ and $\varepsilon_{yy}$ in the defect-free regions, for example in OD 1, are fluctuated about $\pm 1.0\%$; whereas the $\varepsilon_{xy}$ and $\omega_{xy}$ vary $\pm 0.6\%$ and $\pm 0.2°$, respectively. The values of the shear strain and rotation are close to the detectivity (0.5% and 0.2°). The observed average values of $\varepsilon_{xx}$ and $\varepsilon_{yy}$ indicates a very small deformation occurred. It is highly likely related to the local distortions due to the random shifts of the $Pb^{2+}$ cations. The equivalent displacement calculated from $\varepsilon_{xx}$ and $\varepsilon_{yy}$ is about 0.20 to 0.40Å that is similar to the x-ray diffraction results.[18] Zhukov et al. suggested that in a highly ordered PST, the $Pb^{2+}$ cations shift randomly away from the average center about 0.24 to 0.30Å. These nano-scale distortions reduce the crystal symmetry from cubic in the disorder phase or from rhombohedral in the low-temperature ordered phase to a lower one. Therefore, the interface between two different phases can contains local strain.

Several defects are observed on the strain-field maps. According to their characteristic strain fields, we deduce that those are dislocations.[24,25] Such kind of crystal defect is often observed in the highly ordered domains because it acts as a fast diffusion channel during annealing to increase the degree of 1:1 order.[7] Figures 3(a-d) are the line-scan profile along 'X-Y' marked in Fig. 2, in which a dislocation was included. The maximum value of $\varepsilon_{xx}$ and $\varepsilon_{xy}$ and $\omega_{xy}$ are 7.5%, 9.5% and -5.4°, respectively.

In Figs. 2(c) & (d), the APB is recognized on the maps of $\varepsilon_{xy}$ and $\omega_{xy}$. The line scans across the defect-free APB along the mark 'A-B' are shown in Fig. 4(a) and (b), respectively. The sinusoidal-like profiles are observed in both line-scans. This strain region is 6 nm across that is wider than the APB. The maximum values of $\varepsilon_{xy}$ and $\omega_{xy}$ are about 1.5% and 0.9°, respectively, that are smaller than that in dislocations. The values of $\varepsilon_{xy}$ and $\omega_{xy}$ outside APB diminish to the typical ones (~0.5% and 0.2°) in the ordered domains. As the results, such small lattice deformation concentrates inside the APB and only extends to both sides of ordered domains about 1.5 nm (~4 unit cells). Apparently, their influential strain field is shorter than that induced by dislocations. It is worthy to note that this kind of strain cannot be revealed by x-ray or neutron diffraction. However, our observation of strains in APB is consistent with the results of the convergent-beam electron diffraction (CBED) study,[19] in which small departures from the exact general symmetry of the second-order Laue-zone ring were observed in the defect-free APBs. It was concluded that the presence of a strain gives rise to the slight deviation of the symmetry in CBED pattern.

The stress field can be obtained using theory of elasticity.[24,25] At the moment, it is not available because the measurement of elastic property in immediate or highly ordered PST is lacked in the literature and several Brillouin scattering studies show that the elastic constants of disordered PST vary with the degree of cation order.[9,10]

The present results deserve additional theoretical consideration and experimental studies. Strain or displacement field contributes to the parameters relate to the energy gradient, elastic energy and all electromechanical coupling. Those are essential to modeling the structure and growth of the macroscopic domain in ferroelectric/ferroelastic or the nano-polar regions in relaxors.[11,12,14,31] It is also interesting to experimentally investigate the local strain in PST with different degree of order and other order-disorder perovskite oxides.

**Acknowledgements**

We are grateful to Dr. K. Z. Baba-Kishi for supplying the high quality samples. This study was supported by European Commission under Grant No. 29637.

# References

[1] N. M. Shorrocks, R. W. Whatmore, P. C. Osbond, Ferroelectrics **106**, 387 (1990).

[2] P. Muralt, Rep. Prog. Phys. **64**, 1339 (2001).

[3] L. E. Cross, Ferroelectrics **76**, 241 (1987).

[4] Z. G. Ye, Key Eng. Mater. **155/156**, 81 (1998).

[5] N. Setter and L. E. Cross, J. Appl. Phys. **51**, 4356 (1980).

[6] C. G. F. Stenger and A. J. Buggraaf, Phys. Status Solidi A **61**, 275 (1980).

[7] L. A. Bursill, J. L. Peng, H. Qian and N. Setter, Physica B **205**, 305 (1995).

[8] D. Viehland and J. F. Li, J. Appl. Phys. **75**, 1705 (1994).

[9] F. Jiang, J.-H. Ko, S. Lushnikov and S. Kojima, Jpn. J. Appl. Phys. **40**, 5823 (2001).

[10] A. Pietraszko, A. Trzaskowska, A. Hilczer and Z. Tylczynski, Ferroelectric Lett. **34**, 139 (2007).

[11] W. Cao and L. E. Cross, Phys. Rev. B **44**, 5 (1991).

[12] S. Numbu and D. A. Sagala, Phys. Rev. B **50**, 5838 (1994).

[13] Y. Yamada, T. Iwase, K. Fujishiro, Y. Uesu, Y. Yamashita, I. Tomeno and S. Shimanuki, Ferroelectrics **240**, 1629 (2000).

[14] W. Kleemann, Int. J. Mod. Phys. B **7**, 2469 (1993).

[15] I. G. Siny, R. S. Katiyar and A. S. Bhalla, J. Raman Spectrosc. **29**, 385 (1998).

[16] K. Z. Baba-Kishi, G. Cressey and R. J. Cernik, J. Appl. Crystallogr. **25**, 477 (1992).

[17] P. M. Woodward and K. Z. Baba-Kishi, J. Appl. Crystallogr. **35**, 233 (2002).

[18] S. G. Zhukov, V. V. Chernyshev, L. A. Aslanov, S. B. Vakhrushev and H. Schenk, J. Appl. Crystallogr. **28**, 385 (1995).

[19] K. Z. Baba-Kishi and D. J. Barber, J. Appl. Crystallogr. **23**, 43 (1990).

[20] K. Z. Baba-Kishi, C. W. Tai and X. Meng, Phil. Mag. **86**, 5031 (2006).

[21] X. Pan, W. D. Kaplan, M. Rühle and R. E. Newnham, J. Am. Ceram. Soc. **81**, 597 (1998).

[22] M. J. Hÿtch, E. Snoeck and R. Kilaas, Ultramicroscopy **72**, 131 (1998).

[23] M. J. Hÿtch and L. Potez, Phil. Mag. A **76**, 1119 (1997).

[24] M. J. Hÿtch, J.-L. Putaux and J. Thibault, Phil. Mag. **86**, 4641 (2006).

[25] M. J. Hÿtch, J.-L. Putaux and J.-M. Pénisson, Nature **423**, 270 (2003).

[26] K. Z. Baba-Kishi, I. M. Reaney and D. J. Barber, J. Mater. Sci. **25**, 1645 (1990).

[27] C. W. Tai and K. Z. Baba-Kishi, Ferroelectrics **241**, 1645 (2000).

[28] C. W. Tai and K. Z. Baba-Kishi, Textures and Microstructures **35**, 71 (2002).

[29] M. J. Hÿtch and T. Plamann, Ultramicroscopy **87**, 199 (2001).

[30] W. Coene and A. J. E. M. Jansen, Scanning Microsc. Suppl. **6**, 379 (1992).

[31] E. Salje and U. Bismayer, J. Phys.: Condens. Matter **1**, 6967 (1989).

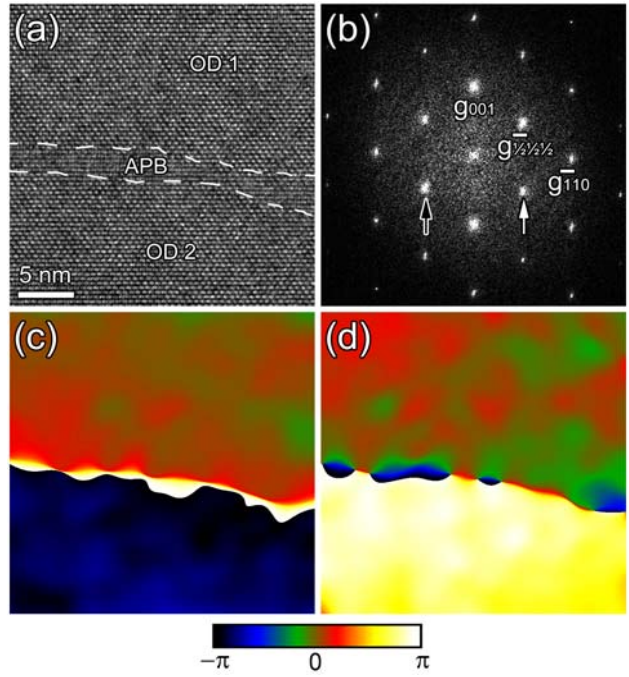

FIG. 1. (a) HRTEM image of two ordered domains (labeled as OD 1 and OD2). The APB is outlined by white dash lines. (b) The power spectrum of the HRTEM image. **g**$_{1/21/2-1/2}$ and **g**$_{-1/21/2-1/2}$ are indicated by the black and white arrows and the corresponding phase images are shown in (c) and (d), respectively.

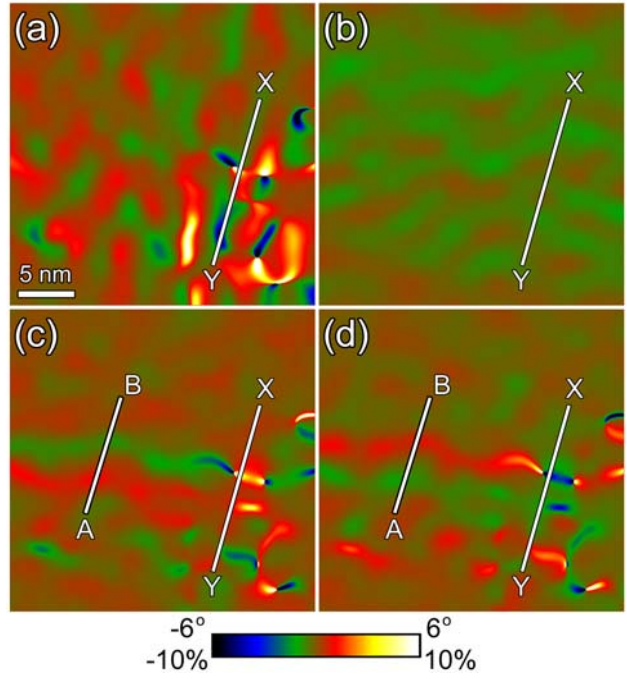

FIG. 2. In-plane strain tensor components (a) $\varepsilon_{xx}$, (b) $\varepsilon_{yy}$ and (c) $\varepsilon_{xy}$ and (d) lattice rotation $\omega_{xy}$. The region of the APB can be seen in (c) and (d).

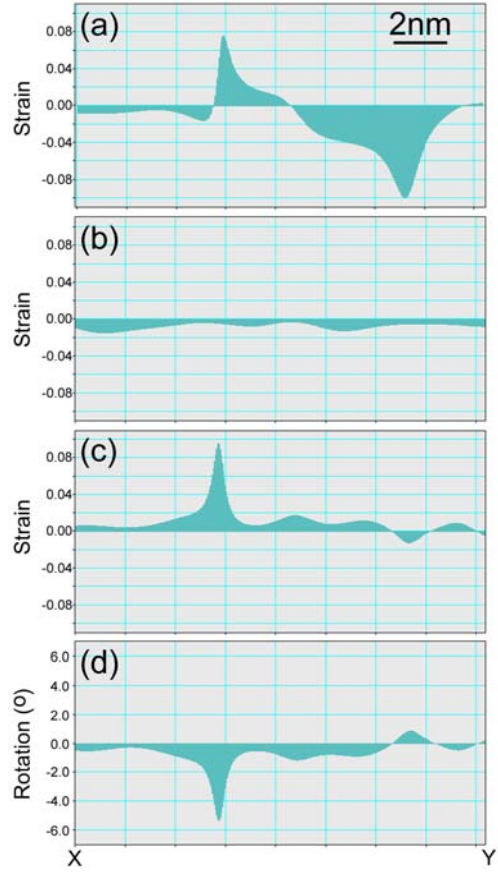

FIG. 3. Profile of (a) $\varepsilon_{xx}$, (b) $\varepsilon_{yy}$, (c) $\varepsilon_{xy}$ and (d) $\omega_{xy}$ along "X-Y" marked in Figs. 2(a-d), respectively.

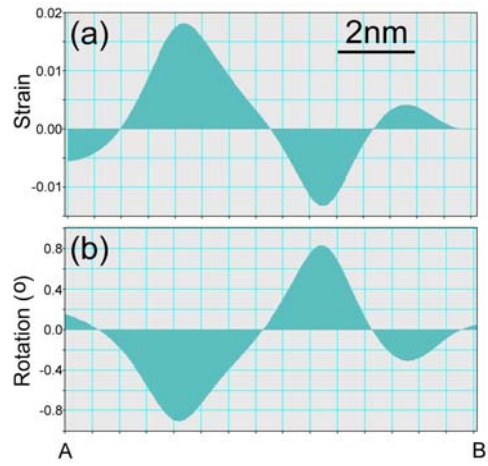

FIG. 4. Profile of (a) $\varepsilon_{xy}$ and (b) $\omega_{xy}$ along "A-B" marked in Figs. 2(c) and (d), respectively.